\documentstyle[preprint,tighten,aps,epsfig]{revtex}
\def\be{\begin{equation}}
\def\ee{\end{equation}}
\def\lsim{\mathrel{\rlap{\lower4pt\hbox{\hskip1pt$\sim$}}
    \raise1pt\hbox{$<$}}}                
\def\slashed{{/}\mskip-10.0mu}
\def\pcircslash{\slashed {p\mskip -5mu ^{^\circ}}}
\begin{document}
\draft
\vskip 2cm

\title{Two-loop additive mass renormalization with clover fermions and Symanzik improved gluons}

\author{A. Skouroupathis, M. Constantinou and H. Panagopoulos}
\address{Department of Physics, University of Cyprus, P.O. Box 20537,
Nicosia CY-1678, Cyprus \\
{\it email: }{\tt php4as01@ucy.ac.cy, phpgmc1@ucy.ac.cy, haris@ucy.ac.cy}}
\vskip 3mm

\date{\today}

\maketitle

\begin{abstract}

We calculate the critical value of the hopping
parameter, $\kappa_c$, in Lattice QCD, up to two
loops in perturbation theory. We employ the Sheikholeslami-Wohlert
(clover) improved action for fermions and the Symanzik improved
gluon action with 4- and 6-link loops.

The quantity which we study is a typical case of a vacuum expectation
value resulting in an additive renormalization; as such, it is
characterized by a power (linear) divergence in the lattice spacing, and
its calculation lies at the limits of applicability of perturbation theory.

Our results are polynomial in $c_{\rm SW}$ (clover parameter) and
cover a wide range of values for the Symanzik coefficients $c_i$. 
The dependence on the number of colors $N$ and the
number of fermion flavors $N_f$ is shown explicitly. In order to
compare our results to non perturbative evaluations of $\kappa_c$
coming from Monte Carlo simulations, we employ an improved
perturbation theory method for improved actions.


\end{abstract}

\newpage


\section{Introduction}
\label{introduction}

In the present work, we calculate the additive 
renormalization of the fermion mass in lattice QCD, using clover
fermions and Symanzik improved gluons. 
The calculation is carried out up to two loops in perturbation
theory and it is directly related to the determination of the critical
value of the hopping parameter, $\kappa_c$.  

The clover fermion action~\cite{SW} (SW) succesfully reduces 
lattice discretization effects and approaches the continuum limit faster.
This justifies the extensive usage of this action in Monte Carlo simulations
in recent years. The coefficient $c_{\rm SW}$ appearing in this
action is a free parameter for the current work and our results will be
given as a polynomial in $c_{\rm SW}$. 

Regarding gluon fields, we employ the Symanzik improved
action~\cite{Sym}, 
which also aims at minimizing finite lattice 
spacing effects. For the coefficients
parameterizing the Symanzik action, we consider several choices of
values which are frequently used in the literature.

The lattice discretization of fermions introduces
some well known difficulties; demanding strict locality and absence
of doublers leads to breaking of 
chiral symmetry. In order to recover this symmetry in the continuum
limit one must set 
the renormalized fermion mass ($m_R$) equal to zero. To achieve
this, the mass parameter $m_\circ$ appearing in the Lagrangian must approach a
critical value $m_c$\,, which is nonzero due to additive
renormalization.

The mass parameter $m_\circ$ is directly related to the hopping
parameter $\kappa$ used in simulations. Its critical value, $\kappa_c$,
corresponds to chiral symmetry restoration: 
\begin{equation}
\kappa_c={1\over 2\,m_c\,a + 8\,r}
\label{kappa1}
\end{equation}
where $a$ is the lattice spacing and $r$ is the Wilson parameter. Using 
Eq.(\ref{kappa1}), the non-renormalized fermion mass is given by:
\begin{equation}
m_B\equiv m_{\rm o}-m_c=\frac{1}{2\,a}\left(\frac{1}{\kappa}
-\frac{1}{\kappa_c}\right)
\label{mB}
\end{equation}
Thus, in order to restore chiral symmetry one must consider the limit
$m_{\rm o} \to m_c$. This fact points to the necessity of an
evaluation of $m_c$\,. 

The perturbative value of $m_c$ is also a necessary ingredient in
higher-loop
calculations of the multiplicative renormalization of operators (see,
e.g., Ref.~\cite{SP}). In
mass independent schemes, such renormalizations are typically defined
and calculated at zero renormalized mass, and this entails setting the
value of the Lagrangian mass equal to $m_c$\,.
  
Previous studies of the hopping parameter and its critical value have appeared
in the literature for Wilson fermions - Wilson gluons~\cite{FP} 
and for clover fermions - Wilson gluons~\cite{PP,CPR}. The 
procedure and notation in our work is the same as in the above references.

Our results for $\kappa_c$ (and consequently for the critical fermion mass)
depend on the number of colors ($N$) and on the number of fermion flavors
($N_f$). Besides that, there is an explicit dependence on the clover
parameter $c_{\rm SW}$ which,  
as mentioned at the beginning, is kept as a free parameter. 
On the other hand, the dependence of the results on the choice of Symanzik coefficients
cannot be given in closed form; instead, we present it in
a list of Tables and Figures.

The rest of the paper is organized as follows: In Sec.~\ref{sec2} we
formulate the problem, define the discretized actions, and describe our 
calculation of the necessary Feynman diagrams. 
Sec.~\ref{sec3} is a presentation of our results. Finally, in Sec.~\ref{sec4} we
apply to our one- and two-loop results an improvement method, proposed by us
\cite{cactus1,cactus2,CPS}. This method resums a certain infinite class of
subdiagrams, to all orders in perturbation theory, leading to an improved
perturbative expansion. We end this section with a comparison of perturbative and
non-perturbative results. Our findings are summarized in Sec.~\ref{sec5}.

\section{Formulation of the problem}
\label{sec2}

We begin with the Wilson formulation of the QCD action on the
lattice, with $N_f$ flavors of degenerate clover (SW)~\cite{SW}
fermions. In standard notation, it reads:

\begin{eqnarray}
S_L =&& S_G + 
\sum_{f}\sum_{x} (4r+m_{\rm o})\bar{\psi}_{f}(x)\psi_f(x)
\nonumber \\
&&- {1\over 2}\sum_{f}\sum_{x,\,\mu}
\bigg{[}\bar{\psi}_{f}(x) \left( r - \gamma_\mu\right)
U_{x,\, x+\mu}\psi_f(x+{\mu}) 
+\bar{\psi}_f(x+{\mu})\left( r + \gamma_\mu\right)U_{x+\mu,\,x}\psi_{f}(x)\bigg{]}\nonumber \\
&&+ {i\over 4}\,c_{\rm SW}\,\sum_{f}\sum_{x,\,\mu,\,\nu} \bar{\psi}_{f}(x)
\sigma_{\mu\nu} {\hat F}_{\mu\nu}(x) \psi_f(x),
\label{latact}
\end{eqnarray}
\begin{eqnarray}
{\rm where:}\qquad {\hat F}_{\mu\nu} &\equiv& {1\over{8}}\,
(Q_{\mu\nu} - Q_{\nu\mu})\\
{\rm and:\qquad} Q_{\mu\nu} &=& U_{x,\, x+\mu}U_{x+\mu,\, x+\mu+\nu}U_{x+\mu+\nu,\, x+\nu}U_{x+\nu,\, x}\nonumber \\
&+& U_{ x,\, x+ \nu}U_{ x+ \nu,\, x+ \nu- \mu}U_{ x+ \nu- \mu,\, x- \mu}U_{ x- \mu,\, x} \nonumber \\
&+& U_{ x,\, x- \mu}U_{ x- \mu,\, x- \mu- \nu}U_{ x- \mu- \nu,\, x- \nu}U_{ x- \nu,\, x}\nonumber \\
&+& U_{ x,\, x- \nu}U_{ x- \nu,\, x- \nu+ \mu}U_{ x- \nu+ \mu,\, x+ \mu}U_{ x+ \mu,\, x}
\label{latact2}
\end{eqnarray}

The clover coefficient $c_{\rm SW}$ is treated here as a free parameter.
Particular  choices of values for $c_{\rm SW}$ have been determined
both perturbatively \cite{SW} and non-perturbatively \cite{Luscher1996}, so as to
minimize ${\cal O}(a)$ effects. The Wilson parameter $r$ is set to $r=1$ henceforth;
$f$ is a flavor index; $\sigma_{\mu\nu} =(i/2) [\gamma_\mu,\,\gamma_\nu]$.
Powers of the lattice spacing $a$ have been omitted and may be
directly reinserted by dimensional counting.

Regarding gluons, we use the Symanzik improved gauge field action, involving
Wilson loops with 4 and 6 links\footnote{$1\times 1$ {\em plaquette},
$1\times 2$ {\em rectangle}, $1\times 2$ {\em chair} (bent rectangle),
and $1\times 1\times 1$ {\em parallelogram} wrapped around an
elementary 3-d cube.}: 
\begin{eqnarray}
S_G&=&\frac{2}{g^2} \Bigl[ c_0 \sum_{\rm plaquette} {\rm Re\,
 Tr\,}\{1-U_{\rm plaquette}\} + c_1 \sum_{\rm rectangle} {\rm Re \, Tr\,}\{1- U_{\rm rectangle}\} \nonumber \\
&&\qquad +c_2 \sum_{\rm chair} {\rm Re\, Tr\,}\{1- U_{\rm chair}\}+c_3 \sum_{\rm parallelogram} {\rm Re \,Tr\,}\{1-
U_{\rm parallelogram}\}\Bigr]
\label{gluonaction}
\end{eqnarray}
($g$ is the bare coupling constant). The lowest order expansion of
this action, leading to the gluon 
propagator, is
\begin{equation}
S_{\rm G}^{(0)} = \frac{1}{2}\int_{-\pi/a}^{\pi/a} \frac{d^4k}{(2\pi)^4}
\sum_{\mu\nu}
A_\mu^a(k)\left[G_{\mu\nu}(k)-\frac{\xi}{\xi-1}\hat{k}_\mu\hat{k}_\nu\right]
A_\nu^a(-k)\,
\label{GluonProp}
\end{equation}
where $\xi$ is the gauge fixing parameter (see Eq.(\ref{GaugeFixing})\,) and:
\begin{eqnarray}
G_{\mu\nu}(k) &=& \hat{k}_\mu\hat{k}_\nu + \sum_\rho \left(
\hat{k}_\rho^2 \delta_{\mu\nu} - \hat{k}_\mu\hat{k}_\rho \delta_{\rho\nu}
\right)  \, d_{\mu\rho} , \qquad \hat{k}_\mu = \frac{2}{a}\sin\frac{ak_\mu}{2}\,, \quad
        \hat{k}^2 = \sum_\mu \hat{k}_\mu^2 
\nonumber\\
d_{\mu\nu}&=&\left(1-\delta_{\mu\nu}\right)
\left[C_0 -
C_1 \, a^2 \hat{k}^2 -  C_2 \, a^2( \hat{k}_\mu^2 + \hat{k}_\nu^2)
\right]
\end{eqnarray}

The coefficients $C_i$ are related to $c_i$ by
\be
C_0 = c_0 + 8 c_1 + 16 c_2 + 8 c_3 \,, \,\,\,
C_1 = c_2 + c_3\,, \,\,\, C_2 = c_1 - c_2 - c_3
\label{ContLimit}
\ee
The Symanzik coefficients must satisfy: $c_0 + 8 c_1 + 16 c_2 + 8 c_3
= 1$, in order to reach the correct classical continuum limit. Aside
from this requirement, the values of $c_i$ can be
chosen arbitrarily; they are normally
tuned in a way as to ensure ${\cal O}(a)$ improvement.

As always in perturbation theory, we must introduce an appropriate
gauge-fixing term to the action; in terms of the gauge field
$Q_\mu(x)$ $\left[U_{x,\,x+\mu}= \exp(i\,g\,Q_\mu(x))\right]$, it reads:
\begin{equation}
S_{gf} = {1\over {1{-}\xi}}\,\sum_{x,\mu , \nu}
\hbox{Tr} \, \bigl\{ \Delta^-_{\mu} Q_{\mu}(x) \Delta^-_{\nu}
Q_{\nu}(x)\bigr\} , \qquad
\Delta^-_{\mu} Q_{\nu}(x) \equiv Q_{\nu}(x - {\hat \mu}) - Q_{\nu}(x).
\label{GaugeFixing}
\end{equation}

Having to compute a gauge invariant quantity, we can, for convenience,
choose to work
either in the Feynman gauge ($\xi=0$) or in the Landau gauge ($\xi=1$).
Covariant gauge fixing produces the following
action for the ghost fields $\omega$ and $\overline\omega$
\begin{eqnarray}
S_{gh} &=& 2 \sum_{x} \sum_{\mu} \hbox{Tr} \,
\Bigl\{(\Delta^+_{\mu}\omega(x))^{\dagger} \Bigl( \Delta^+_{\mu}\omega(x) +
i g \left[Q_{\mu}(x),
\omega(x)\right] + \frac{1}{2}
i g \left[Q_{\mu}(x), \Delta^+_{\mu}\omega(x) \right] \nonumber\\
 & & \quad - \frac{1}{12}
g^2 \left[Q_{\mu}(x), \left[ Q_{\mu}(x),
\Delta^+_{\mu}\omega(x)\right]\right] + \cdots \Bigr)\Bigr\}, \qquad
\Delta^+_{\mu}\omega(x) \equiv \omega(x + {\hat \mu}) - \omega(x).
\label{GhostAction}
\end{eqnarray}
Finally, the change of integration variables from links to vector
fields yields a Jacobian that can be rewritten as
the usual measure term $S_m$ in the action:
\begin{equation}
S_{m} = \frac{1}{12} N g^2 \sum_{x} \sum_{\mu} \hbox{Tr} \,
\{Q_{\mu}(x) Q_{\mu}(x)\} + \cdots
\end{equation}
In $S_{gh}$ and $S_m$ we have written out only
terms relevant to our computation.
The full action is: $S = S_L + S_{gf} + S_{gh} + S_m.$

The bare fermion mass $m_B$ must be set to zero for
chiral invariance in the classical continuum limit. 
Terms proportional to $r$ in the action, as well as the clover terms,
break chiral invariance. They vanish in the classical continuum
limit; at the quantum level, they induce nonvanishing,
flavor-independent fermion mass corrections.
Numerical simulation algorithms usually employ the hopping parameter,
\begin{equation}
\kappa\equiv{1\over 2\,m_{\rm o}\,a + 8\,r}
\label{kappaDef}
\end{equation}
as an adjustable input. Its critical value, at which chiral symmetry
is restored, is thus $1/8r$ classically, but gets shifted by quantum
effects.

The renormalized mass can be calculated in textbook fashion from the
fermion self--energy. Denoting by $\Sigma^L(p,m_{\rm o},g)$ the truncated,
one particle irreducible fermion two-point function, we have for the
fermion propagator:
\begin{eqnarray}
S(p)&=& \left[ i \,\pcircslash + m(p)- \Sigma^L(p,m_{\rm o},g)\right]^{-1}\\
{\rm where:}\qquad \pcircslash &=& {1\over a}\,\sum_\mu\gamma_\mu  \,\sin(ap^\mu), \quad m(p) = m_{\rm o} + {2r\over a} \sum_\mu \sin^2(ap^\mu/2)\nonumber
\end{eqnarray}

To restore the explicit breaking of chiral invariance, we require that the renormalized mass vanish:
\begin{equation}
S^{-1}(0)\Big|_{\displaystyle m_{\rm o} \to m_c} = 0 \qquad\Longrightarrow\qquad m_c = \Sigma^L(0,m_c\,,g)
\end{equation}
The above is a recursive equation for $m_c$, which can be solved order
by order in perturbation theory.

\bigskip
We denote by $dm$ the additive mass renormalization of $m_\circ$\,:
$m_B = m_\circ - dm$. In order to obtain a zero renormalized mass, we
must require $m_B \to 0$, and thus $m_\circ \to dm$. Consequently,
\begin{equation}
m_c = dm= dm_{\rm (1-loop)}+dm_{\rm (2-loop)}
\label{Totaldm}
\end{equation}
At tree level, $m_c=0$.

Two diagrams contribute to $dm_{\rm (1-loop)}$, shown in Fig. 1. In these
diagrams, the fermion mass must be set to its tree level value, 
$m_{\rm o}\to 0$.

\bigskip
\hskip4.0cm\psfig{figure=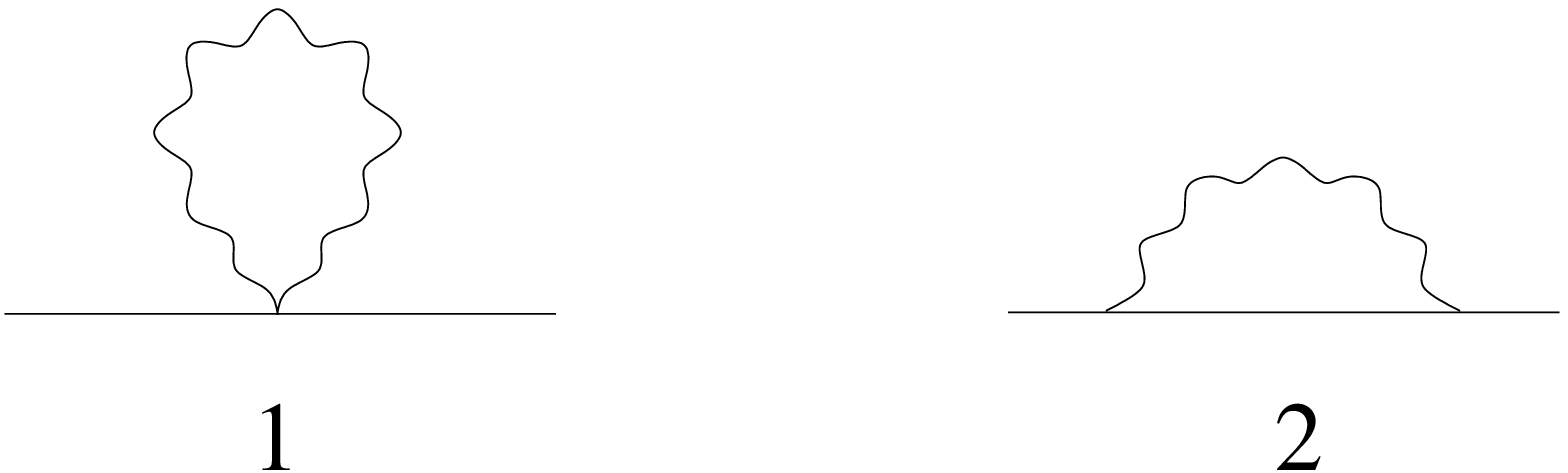,height=2truecm}\hskip1.0cm

\begin{center}
\begin{minipage}{9.3cm}\noindent
\begin{center}
{\small {\bf Fig. 1.} One-loop diagrams contributing to $dm_{\rm (1-loop)}$.
Wavy (solid) lines represent gluons (fermions).}
\end{center}
\end{minipage}
\end{center}

\smallskip
The quantity $dm_{\rm (2-loop)}$ receives contributions from
a total of 26 diagrams, shown in
Fig. 2. Genuine two-loop diagrams must again be evaluated at
$m_{\rm o}\to 0$; in addition, one must include to this order the one-loop diagram
containing an ${\cal O}(g^2)$ mass counterterm (diagram 23).

Certain sets of diagrams, corresponding to one-loop renormalization of 
propagators, must be evaluated together in
order to obtain an infrared convergent result:
These are diagrams 7+8+9+10+11, 12+13, 14+15+16+17+18, 19+20, 21+22+23.

\hskip2.0cm\psfig{figure=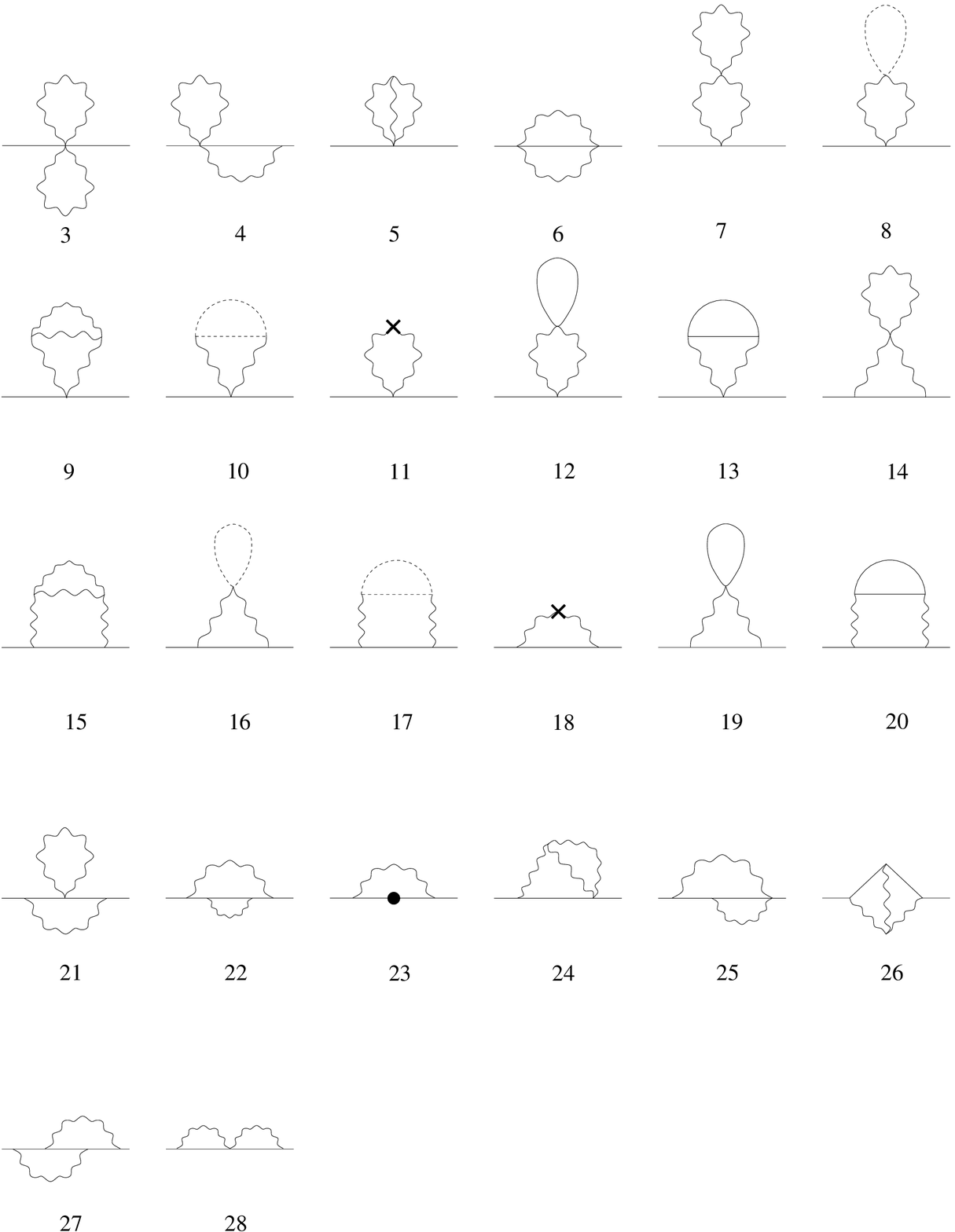,height =14truecm}\hskip1.0cm
\bigskip
\nobreak

\noindent
{\small {\bf Fig. 2.} Two-loop diagrams contributing to $dm_{\rm (2-loop)}$.
Wavy (solid, dotted) lines represent gluons (fermions,
ghosts). Crosses denote vertices stemming from the measure part of the
action; a solid circle is a fermion mass counterterm.}

\section{Computation and Results}
\label{sec3}

Given that the dependence of $m_c$ on the Symanzik coefficients $c_i$ cannot
be expressed in closed form, we chose certain sets of values for $c_i$
, presented in Table I, which are in common
use~\cite{LWactions,Iwasaki,Symanzik,LWactions2,Alford,Takaishi}:
Plaquette, Symanzik (tree level improved), Tadpole Improved L\"uscher-Weisz
(TILW), Iwasaki and DBW2.
Actually, since the gluon
propagator contains only the combinations $C_1$ and $C_2$\, (Eq.(\ref{ContLimit})), all
results for $m_c$ can be recast in terms of $C_1$, $C_2$ and one additional
parameter, say, $c_2$;
in this case the dependence on $c_2$ (at fixed $C_1$, $C_2$) is
polynomial of second degree.

The contribution $dm_l$ of the $l^{{\rm th}}$ one-loop diagram to $dm$, can be
expressed as:
\begin{equation}
dm_l=\frac{(N^2-1)}{N}\,g^2 \cdot \sum_{i=0}^{2}c_{{\rm
    SW}}^i\,\varepsilon^{(i)}_{l}  
\label{1loopContribution}
\end{equation}
where $\varepsilon^{(i)}_{l}$ are numerical one-loop integrals whose values
depend on $C_1$, $C_2$. The 
dependence on $c_{{\rm SW}}$ is seen to be polynomial of
degree 2 ($i=0,\,1,\,2$). 

The contribution to $dm$ from two-loop diagrams that do
not contain closed fermion loops, can be written in the form
\begin{equation}
dm_l=\frac{(N^2-1)}{N^2}\,g^4 \cdot \sum_{i,j,k} c_{{\rm SW}}^i\,N^j\,c_2^k\,e^{(i,j,k)}_{l}
\label{2loopContribution1}
\end{equation}

\noindent
where the index $l$ runs over all contributing diagrams, $j=0,2$ and
$k=0,\,1,\,2$ (since up to two vertices from the gluon action may be
present in a Feynman diagram). The dependence on $c_{{\rm SW}}$ is now
polynomial of
degree 4 ($i=0,\cdots,\,4$). The coefficients $e^{(i,j,k)}_{l}$ (as
well as $\tilde{e}^{(i)}_{l}$ of Eq.(\ref{2loopContribution2}) below) are
two-loop numerical integrals; once again, they depend on $C_1$, $C_2$.
Finally, the contribution to $dm$ from two-loop diagrams
containing a closed fermion loop, can be expressed as
\begin{equation}
dm_l=\frac{(N^2-1)}{N}\,N_f\,g^4 \cdot
\sum_{i=0}^{4}c^{i}_{\rm SW}\,\tilde{e}^{(i)}_{l}
\label{2loopContribution2}
\end{equation}
where the index $l$ runs over diagrams 12-13, 19-20.
Summing up the contributions of all diagrams, $dm$ assumes the form
\begin{eqnarray}
dm=\sum_{l}dm_l&=&\frac{(N^2-1)}{N}\,g^2 \cdot \sum_{i}c_{{\rm SW}}^i\,\varepsilon^{(i)}
+\frac{(N^2-1)}{N^2}\,g^4 \cdot \sum_{i,j,k}c_{{\rm SW}}^i\,N^j\,c_2^k\,e^{(i,j,k)} \nonumber \\
&&+\frac{(N^2-1)}{N}\,N_f\,g^4 \cdot \sum_i c^{i}_{\rm SW}\,\tilde{e}^{(i)}
\label{TotalContribution}
\end{eqnarray}
In the above, $\varepsilon^{(i)}$, $e^{(i,j,k)}$, $\tilde{e}^{(i)}$
are the sums over all contributing diagrams of the quantities:
$\varepsilon^{(i)}_{l}$, $e^{(i,j,k)}_{l}$, $\tilde{e}^{(i)}_{l}$,
respectively 
(cf. Eqs.(\ref{1loopContribution},\ref{2loopContribution1},\ref{2loopContribution2})\,).

The coefficients $\varepsilon^{(i)}$ lead to the total contribution of
one-loop diagrams. Their values are listed in Table \ref{tab2}, for the ten
sets of $c_i$ values shown in Table \ref{tab1}.
Similarly, results for the coefficients
$e^{(i,j,k)}$ and $\tilde{e}^{(i)}$ corresponding to the total
contribution of two-loop diagrams, are presented in
Tables \ref{tab3}-\ref{tab7}.    

In order to enable cross-checks and comparisons, numerical
per-diagram values of the constants $\varepsilon^{(i)}_{l}$, 
$e^{(i,j,k)}_{l}$ and $\tilde{e}^{(i)}_{l}$ are presented in Tables
VIII-XII, for the case of the Iwasaki action. For economy of space,
several vanishing contributions to these constants have simply been
omitted. A similar breakdown for
other actions can be obtained from the authors upon request.

The total contribution of one-loop diagrams, for $N=3$ can
be written as a function of the clover parameter $c_{{\rm SW}}$.
In the case of the Plaquette, Iwasaki, and DBW2 actions, we find, respectively:
\begin{eqnarray}
dm_{\rm (1-loop)}^{\rm Plaquette}&=&g^2\,\Big(-0.434285489(1)+0.1159547570(3)\,c_{{\rm SW}}+0.0482553833(1)\,c_{{\rm SW}}^2\Big)
\label{dm1loopPlaquette}\\
dm_{\rm (1-loop)}^{\rm Iwasaki}&=&g^2\,\Big(-0.2201449497(1)+0.0761203698(3)\,c_{{\rm SW}}+0.0262264231(1)\,c_{{\rm SW}}^2\Big)
\label{dm1loopIwasaki}\\
dm_{\rm (1-loop)}^{\rm DBW2}&=&g^2\,\Big(-0.0972070995(5) + 0.0421775310(1)\,c_{{\rm SW}} + 0.01141359801(1)\,c_{{\rm SW}}^2\Big)
\label{dm1loopDBW2}
\end{eqnarray}
A similar process can be followed for two-loop diagrams. In
this case, we set $N=3$, $c_2=0$ and we use three
different values for the flavor number: $N_f=0,\,2,\,3$. Thus, for
the Plaquette, Iwasaki and DBW2 actions, the total contribution is, respectively:
\begin{eqnarray}
N_f=0:&& \quad
dm_{\rm (2-loop)}^{\rm Plaquette}=g^4\,\Big(-0.1255626(2)+0.0203001(2)\,c_{{\rm SW}}+0.00108420(7)\,c_{{\rm SW}}^2 \nonumber \\
&& \hspace{3.6cm} -\,0.00116538(2)\,c_{{\rm SW}}^3-0.0000996725(1)\,c_{{\rm SW}}^4\Big) \\
N_f=2:&& \quad
dm_{\rm (2-loop)}^{\rm Plaquette}=g^4\,\Big(-0.1192361(2)+0.0173870(2)\,c_{{\rm SW}}+0.00836498(8)\,c_{{\rm SW}}^2 \nonumber \\
&& \hspace{3.6cm} -\,0.00485727(3)\,c_{{\rm SW}}^3-0.0011561947(4)\,c_{{\rm SW}}^4\Big) \\
N_f=3:&& \quad
dm_{\rm (2-loop)}^{\rm Plaquette}=g^4\,\Big(-0.1160729(2)+0.0159305(2)\,c_{{\rm SW}}+0.0120054(1)\,c_{{\rm SW}}^2 \nonumber \\
&& \hspace{3.6cm} -\,0.00670321(3)\,c_{{\rm SW}}^3-0.0016844558(6)\,c_{{\rm SW}}^4\Big)
\label{dm2loopPlaquette}
\end{eqnarray}
\begin{eqnarray}
N_f=0:&& \quad
dm_{\rm (2-loop)}^{\rm Iwasaki}=g^4\,\Big(-0.0099523(2)-0.0024304(5)\,c_{{\rm SW}}-0.00232855(4)\,c_{{\rm SW}}^2 \nonumber \\
&& \hspace{3.5cm} -\,0.00032100(2)\,c_{{\rm SW}}^3-0.0000419365(1)\,c_{{\rm SW}}^4\Big)\\
N_f=2:&& \quad
dm_{\rm (2-loop)}^{\rm Iwasaki}=g^4\,\Big(-0.0076299(2)-0.0040731(5)\,c_{{\rm SW}}+0.00102758(6)\,c_{{\rm SW}}^2 \nonumber \\
&& \hspace{3.5cm} -\,0.00242924(3)\,c_{{\rm SW}}^3-0.000457690(2)\,c_{{\rm SW}}^4\Big)\\
N_f=3:&& \quad
dm_{\rm (2-loop)}^{\rm Iwasaki}=g^4\,\Big(-0.0064687(2)-0.0048944(5)\,c_{{\rm SW}}+0.00270565(7)\,c_{{\rm SW}}^2 \nonumber \\
&& \hspace{3.5cm} -\,0.00348335(3)\,c_{{\rm SW}}^3-0.000665567(2)\,c_{{\rm SW}}^4\Big)
\label{dm2loopIwasaki}
\end{eqnarray}
\begin{eqnarray}
N_f=0:&& \quad
dm_{\rm (2-loop)}^{\rm DBW2}=g^4\,\Big(+0.005099(2)-0.0053903(7)\,c_{{\rm SW}}-0.0011157(1)\,c_{{\rm SW}}^2 \nonumber \\
&& \hspace{3.5cm} -\,0.00004482(2)\,c_{{\rm SW}}^3-0.0000111470(2)\,c_{{\rm SW}}^4\Big)\\
N_f=2:&& \quad
dm_{\rm (2-loop)}^{\rm DBW2}=g^4\,\Big(+0.005944(2) - 0.0061840(7)\,c_{{\rm SW}}+0.0002046(2)\,c_{{\rm SW}}^2 \nonumber \\
&& \hspace{3.5cm} -\,0.0010177(3)\,c_{{\rm SW}}^3 -0.000125065(3)\,c_{{\rm SW}}^4\Big)\\
N_f=3:&& \quad
dm_{\rm (2-loop)}^{\rm DBW2}=g^4\,\Big(+0.006366(2)-0.0065809(7)\,c_{{\rm SW}}+0.0008648(2)\,c_{{\rm SW}}^2 \nonumber \\
&& \hspace{3.5cm} -\,0.0015042(4)\,c_{{\rm SW}}^3 -0.000182023(5)\,c_{{\rm SW}}^4\Big)
\label{dm2loopDBW2}
\end{eqnarray}

In Figs. 3, 4, and 5 we present the values of $dm_{\rm (2-loop)}$
for $N_f=0,\,2,\,3$, respectively; the results are shown for all
choices of Symanzik actions which we have considered, as a function of
$c_{\rm SW}\,(N=3,\,c_2=0)$. In all cases, the dependence on 
$c_{\rm SW}$ is rather mild. One observes that $dm_{\rm (2-loop)}$ is
significantly smaller for all improved actions, as compared to
the plaquette action; in particular, in the case of DBW2, $dm_{\rm (2-loop)}$ is closest to
zero and it vanishes exactly around $c_{\rm SW}=1$. 

Another feature of these results is that they change
only slightly with $N_f$\,, especially in the range $c_{\rm SW} <
1.5$\,. This is due to the small contributions of diagrams with closed
fermion loops (diagrams 12, 13, 19, 20). By the same token, in the
case of nondegenerate flavors, $dm_{\rm (2-loop)}$ is expected to
depend only weakly on the mass of the virtual fermion. 

\begin{center}
\psfig{file=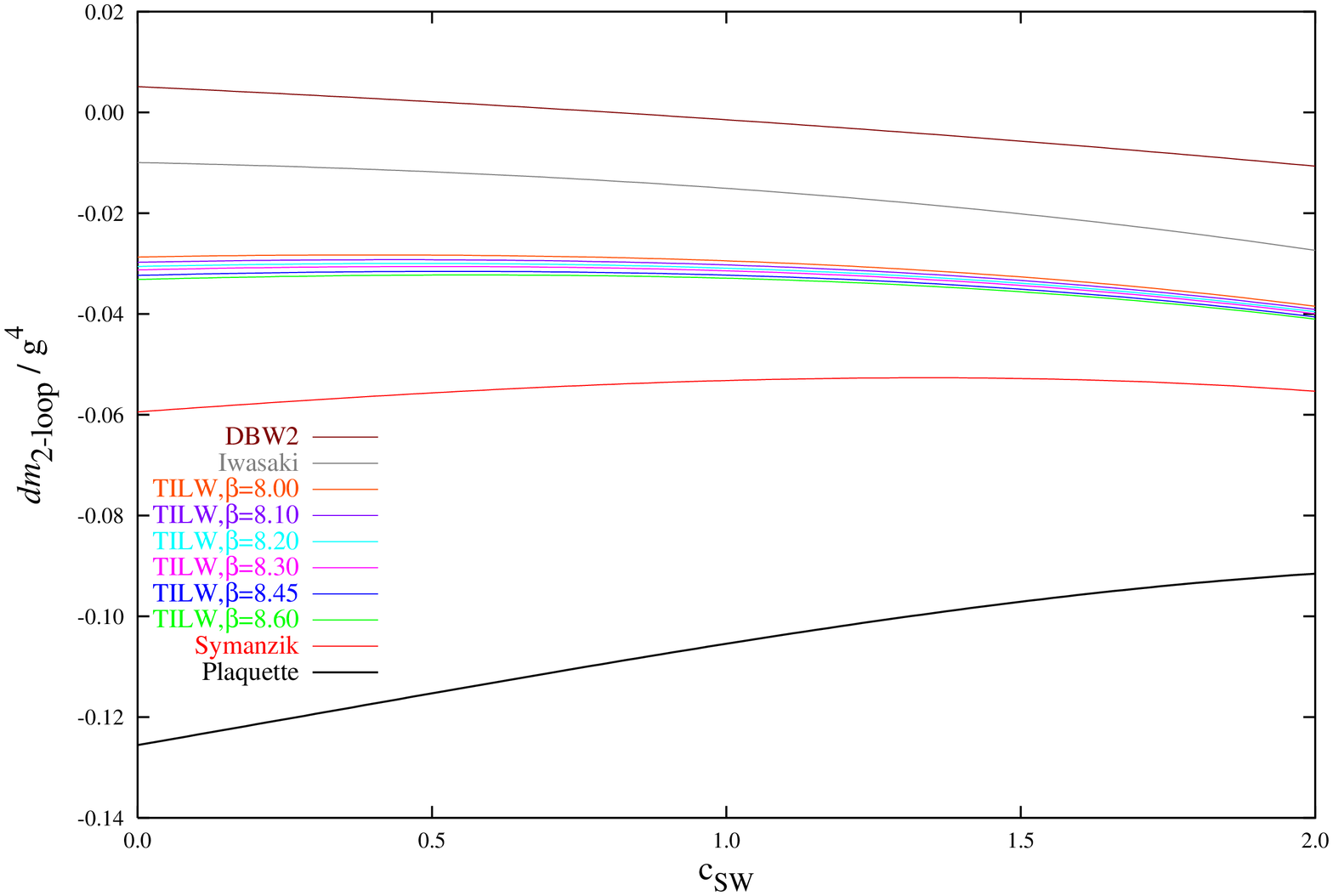,scale=0.55,angle=-90}
\vskip 3mm
\noindent
{\small {\bf Fig. 3.} Total contribution of two-loop diagrams, for
  $N=3$, $N_f=0$ and $c_2=0$. Legends appear in the same top-to-bottom
order as the corresponding lines.}
\vskip -1mm \psfig{file=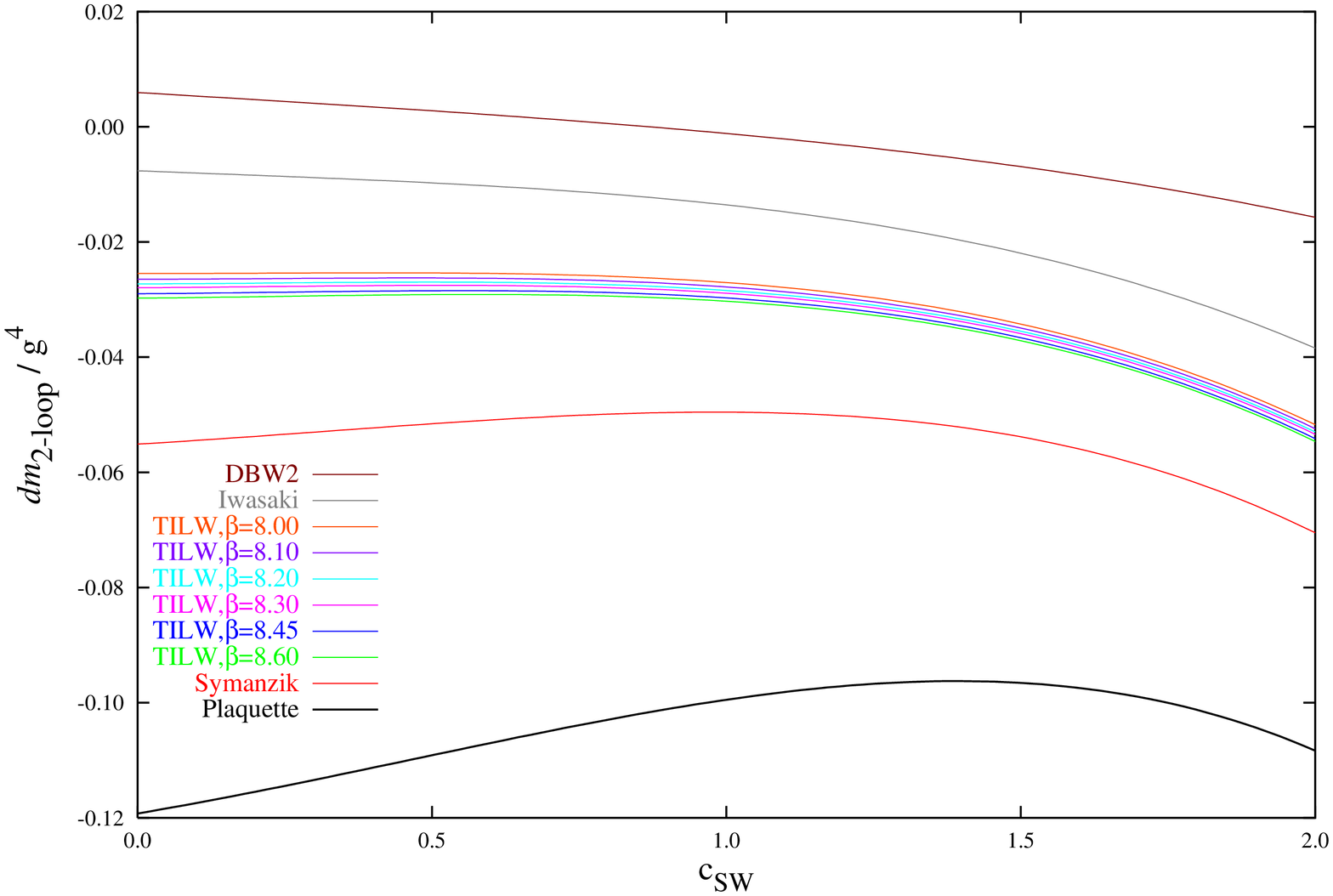,scale=0.55,angle=-90}
\vskip 3mm
\noindent
{\small {\bf Fig. 4.} Total contribution of two-loop diagrams, for
  $N=3$, $N_f=2$ and $c_2=0$. Legends appear in the same top-to-bottom
order as the corresponding lines.
}
\end{center}

\begin{center}
\psfig{file=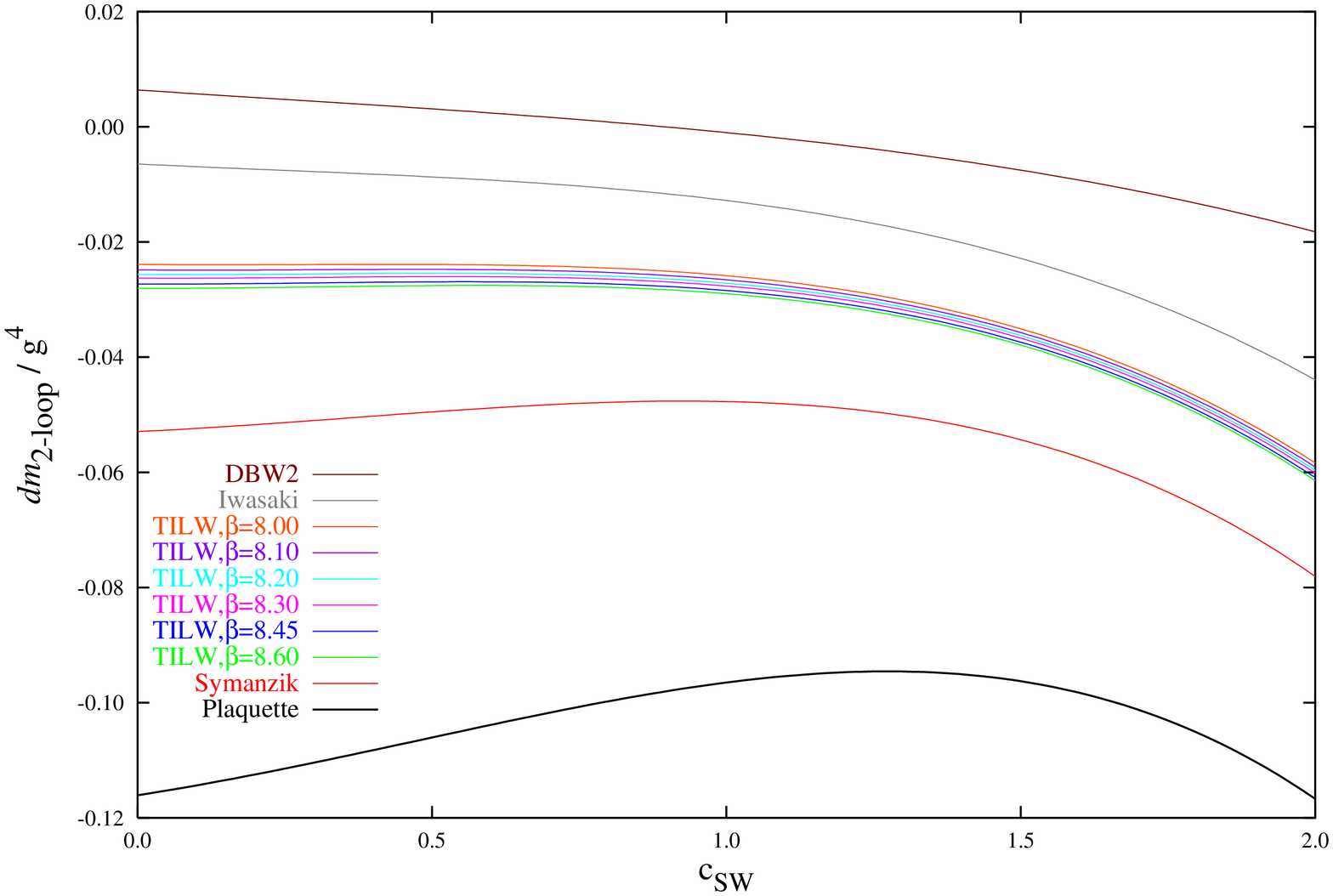,scale=0.55,angle=-90}
\vskip 3mm
\noindent
{\small {\bf Fig. 5.} Total contribution of two-loop diagrams, for
  $N=3$, $N_f=3$ and $c_2=0$. Legends appear in the same top-to-bottom
order as the corresponding lines.}
\end{center}

\section{Improved Perturbation Theory}
\label{sec4}

We now apply our method of improving perturbation theory
\cite{cactus1,cactus2,CPS}, 
based on resummation of an infinite subset of tadpole diagrams, termed
``cactus'' diagrams. In Ref. \cite{CPS} we show how this procedure can
be applied to any action of the type we are considering here, and it
provides a simple, gauge invariant way of dressing, to all orders, perturbative results
at any given order (such as the one- and two-loop results of the present
calculation). Some alternative ways of improving perturbation theory
have been proposed in Refs. \cite{Parisi-81,L-M-93}. In a nutshell,
our procedure involves replacing the 
original values of the Symanzik and clover coefficients by improved
values, which are explicitly computed in \cite{CPS}. Applying at first
this method to one-loop diagrams, the improved
(``dressed'') value $dm^{\rm dr}$ of the critical mass
$(N=3,\,c_2=0)$ can be
written as:
\begin{equation}
dm^{\rm dr}_{\rm (1-loop)}=\sum_{i=0}^{2}\varepsilon^{(i)}_{dr}\,c_{{\rm
SW}}^i \label{dm1loopDressed}
\end{equation}
In comparing with $\varepsilon^{(i)}$ of Eq.
(\ref{TotalContribution}), the quantity $\varepsilon^{(i)}_{dr}$ is
the result of one-loop Feynman diagrams with dressed values for the
Symanzik parameters, and it has already been multiplied by
$g^2\left(N^2-1\right)/N$. The dependence of $\varepsilon^{(i)}_{dr}$
on $g$ is quite complicated now, and cannot be given in closed form;
instead $\varepsilon^{(i)}_{dr}$ must be computed numerically for
particular choices of $g$.
 Listed in Table
\ref{tab13} are the results for $\varepsilon^{(i)}_{dr}$ along
with the value of $\beta=2N/g^2$ corresponding to each one of the 16
actions used in this calculation.

An attractive feature of this improvement procedure is that it can be
applied also to higher loop perturbative results, with due care to
avoid double counting of the cactus diagrams which were already
included at one loop. Ideally, of course, one loop improvement should
already be adequate enough, so as to obviate the need to consider
higher loops; indeed, we find this to be the case and, consequently,
we limit our discussion of two-loop improvement to only the plaquette
action $(\beta=5.29,\,N=3,\,N_f=2)$, the Iwasaki action
$(\beta=1.95,\,N=3,\,N_f=2)$ and the DBW2 action
$(\beta=0.87\ {\rm and}\ \beta=1.04,\,N=3,\,N_f=2)$. Using these values, the contribution to  
$dm^{\rm dr}_{\rm (2-loop)}$ is a polynomial in $c_{\,{\rm SW}}$:
\begin{eqnarray}
dm^{\rm dr}_{\rm (2-loop),\,plaquette}&=& -0.77398(8)+0.16330(4)\,c_{{\rm SW}}+ 0.06224534(1)\,c_{{\rm SW}}^2 \nonumber \\
&& -0.0044006(9)\,c_{{\rm SW}}^3-0.00073780(6)\,c_{{\rm SW}}^4
\label{dm2loopDressedPlaquette}\\[2ex]
dm^{\rm dr}_{\rm (2-loop),\,Iwasaki}&=&-0.0813302(9)+0.043030(3)\,c_{{\rm
SW}}+0.0308196(2)\,c_{{\rm SW}}^2 \nonumber \\
&& -0.00767090(8)\,c_{{\rm SW}}^3-0.001160923(1)\,c_{{\rm SW}}^4
\label{dm2loopDressedIwasaki}\\[2ex]
dm^{\rm dr}_{\rm (2-loop),\,DBW2(\beta=0.87)}&=&-0.044906(1)+0.029449(4)\,c_{{\rm
SW}}+0.0239522(2)\,c_{{\rm SW}}^2 \nonumber \\
&& -0.0082231(1)\,c_{{\rm SW}}^3 -0.001218955(4)\,c_{{\rm SW}}^4
\label{dm2loopDressedDBW2_0.87}\\[2ex]
dm^{\rm dr}_{\rm (2-loop),\,DBW2(\beta=1.04)}&=&-0.031260(1)+0.021793(2)\,c_{{\rm
SW}}+0.0188027(2)\,c_{{\rm SW}}^2 \nonumber \\
&& -0.00705284(9)\,c_{{\rm SW}}^3 -0.001055657(1)\,c_{{\rm SW}}^4
\label{dm2loopDressedDBW2_1.04}
\end{eqnarray}

The comparison between the total dressed contribution
$dm^{\rm dr}=dm^{\rm dr}_{\rm (1-loop)}+dm^{\rm dr}_{\rm (2-loop)}$
and the unimproved 
contribution, $dm$, for the plaquette action is 
exhibited in Fig. 6, as a function of $c_{\rm SW}$. Similarly, $dm^{\rm dr}$
for the Iwasaki and the DBW2 actions is shown in Fig. 7 and Fig. 8,
respectively.  

\bigskip
\begin{center}
\psfig{file=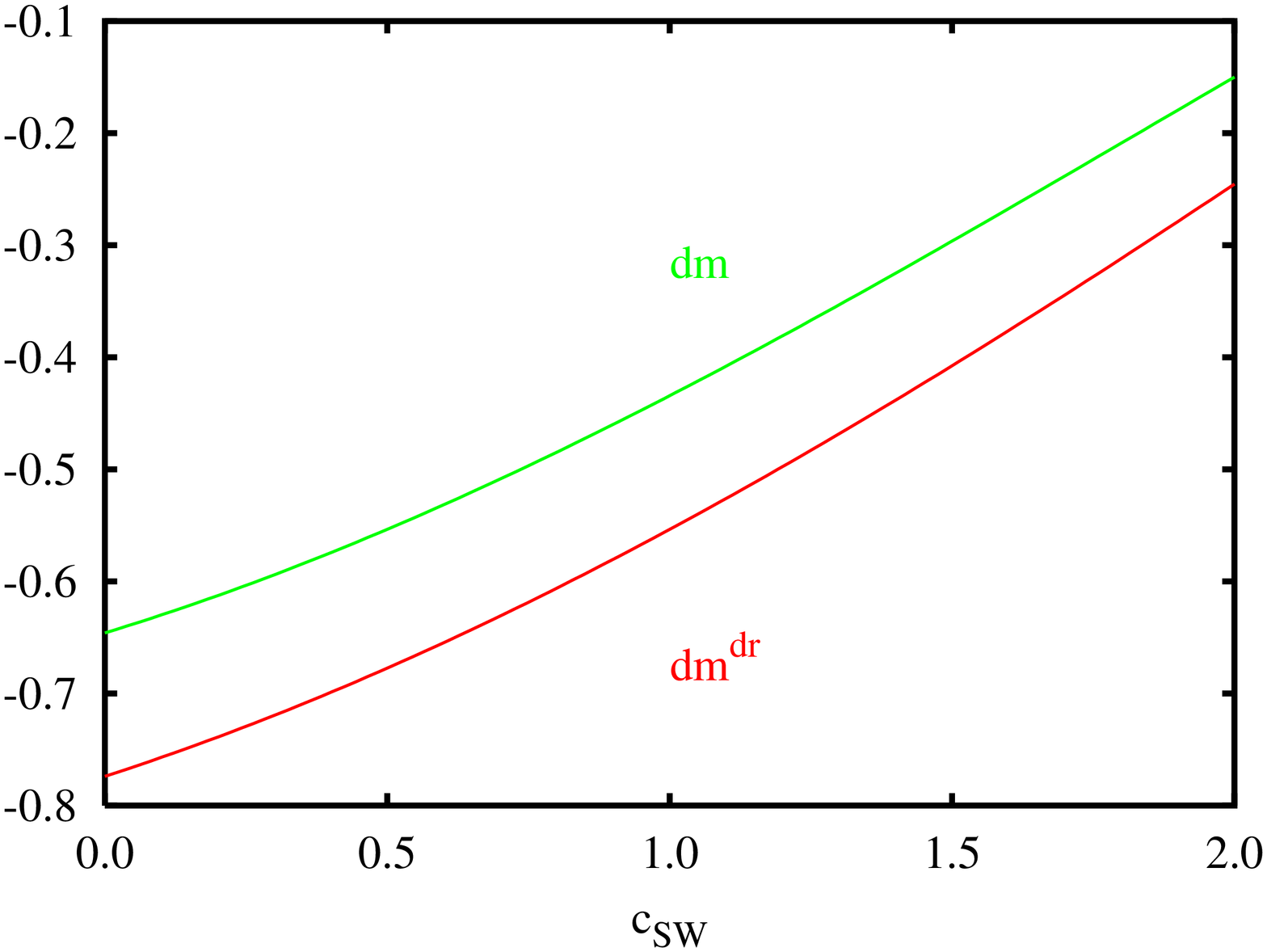,scale=0.5,angle=-90}

\small{{\bf Fig. 6.} Improved and unimproved values of $dm$ up to two
loops, as a function of $c_{\rm SW}$, for the plaquette action
($\beta=5.29$, $N=3$, $N_f=2$).}
\end{center}

\begin{center}
\psfig{file=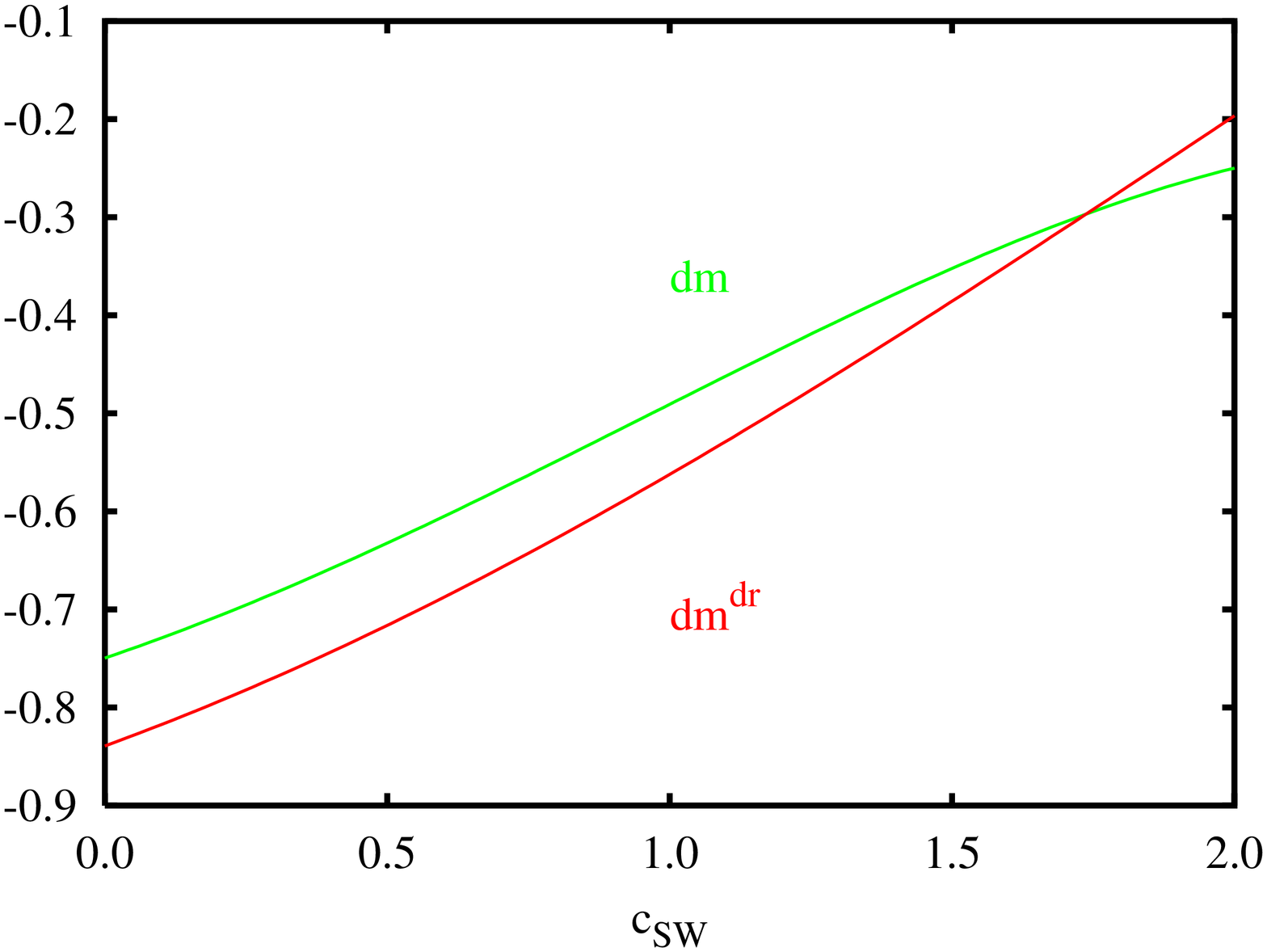,scale=0.5,angle=-90}

\small{{\bf Fig. 7.} Improved and unimproved values of $dm$ up to two
loops, as a function of $c_{\rm SW}$, for the Iwasaki action
($\beta=1.95$, $N=3$, $N_f=2$).}
\end{center}

\begin{center}
\psfig{file=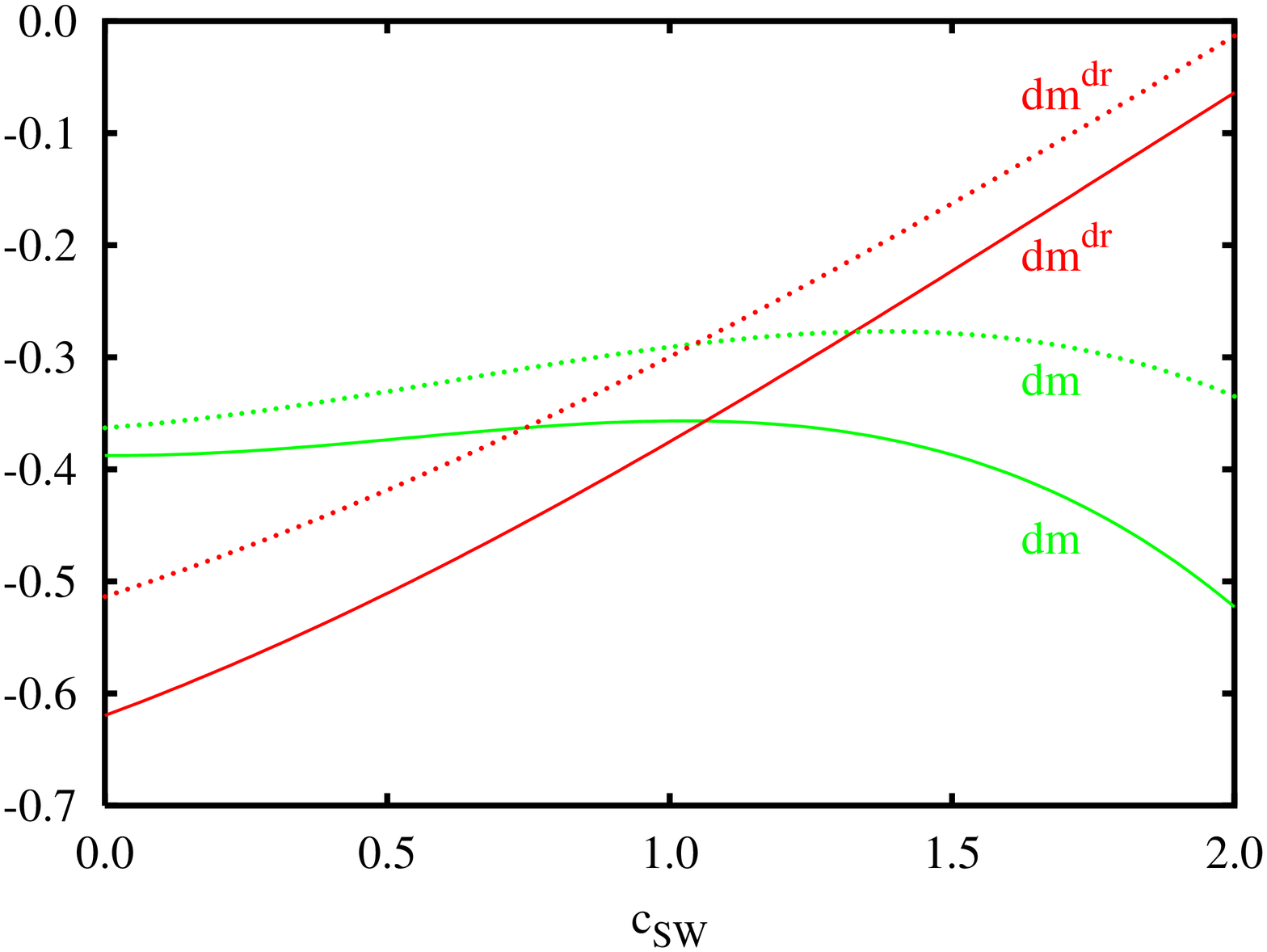,scale=0.5,angle=-90}

\small{{\bf Fig. 8.} Improved and unimproved values of $dm$ up to two
loops, as a function of $c_{\rm SW}$, for the DBW2 action
($N=3$, $N_f=2$). We set $\beta=0.87$ (solid lines) and $\beta=1.04$
(dotted lines).}
\end{center}

Finally, in Table \ref{tab14}, we present a comparison of dressed and
undressed results, for some commonly used values of $\beta$, $N_f$,
$c_{\rm SW}$, and we also compare with available non perturbative estimates for $\kappa_c$ 
\cite{Luscher1996,Bowler,UKQCD,Sommer,Khan}. We observe that improved
perturbation theory, applied to one-loop 
results, already leads to a much better agreement with the non
perturbative estimates.

\section{Discussion}
\label{sec5}

To recapitulate, in this paper we have calculated the critical mass
$m_c$, and the associated critical hopping parameter $\kappa_c$, up to two
loops in perturbation theory, using the clover action for fermions and the Symanzik improved
gluon action with 4- and 6-link loops. The perturbative value of $m_c$
is a necessary ingredient in the higher-loop renormalization of operators, in
mass independent schemes: Such renormalizations are typically defined
and calculated at vanishing renormalized mass, which amounts to setting the
Lagrangian mass equal to $m_c$.

In our calculations, we have chosen for the Symanzik coefficients
$c_i$ a wide range of values, which are most commonly used in
numerical simulations. The dependence of our results on the number of
colors $N$ and the 
number of fermion flavors $N_f$ is shown explicitly. The dependence on
the clover parameter $c_{\rm SW}$ is in the form of a fourth degree
polynomial whose coefficients we compute explicitly; it is expected,
of course, that the most relevant values for $c_{\rm SW}$ are those
optimized for ${\cal O}(a)$ improvement, either at
tree level ($c_{\rm SW}=1$), or at one loop ~\cite{SW}, or
non-perturbatively~\cite{Luscher1996}.

Since $m_c$ is gauge invariant, we chose to calculate it in the
Feynman gauge. The propagator appearing in Feynman diagrams is the
inverse of a nondiagonal matrix; while this inverse can be written
down explicitly, it is more convenient, and more efficient in terms of
CPU time, to perform the inversion numerically. Integrations over loop momenta
were performed as momentum sums on lattices of finite size $L$, where
typically $L\lsim 40$; extrapolation to $L\to\infty$ introduces a
systematic error, which we estimate quite accurately.

Our results for $m_c$ are significantly closer to zero in the case of
Symanzik improved actions, as compared to the plaquette action. In
particular, the DBW2 action stands out among the rest, in that $m_c$
vanishes exactly for a value of $c_{\rm SW}$ around 1. Thus,
improved actions seem to bring us quite near the point of chiral
symmetry restoration.
The dependence of $m_c$ on the number of flavors is seen to be very
mild. This fact would also suggest that, in the
case of nondegenerate flavors, $m_c$ should 
depend only weakly on the mass of the virtual fermion. 

Finally, we have made some comparisons among perturbative and
non-perturbative results for $\kappa_c$. While these are expected to differ
for a power divergent additive renormalization, such as the quantity
under study, we nevertheless find a reasonable agreement. This
agreement is further enhanced upon using an improved perturbative
scheme, which
entails resumming, to all orders in the coupling constant, a dominant
subclass of tadpole diagrams. The method, originally proposed for the
Plaquette action (see Ref. \cite{cactus1}), was extended in
Ref. \cite{CPS} to encompass all possible gluon actions made of closed 
Wilson loops, and can be applied at any given order in perturbation
theory. As would be desirable, one-loop improvement is seen to be already
adequate to give a 
reasonable agreement among perturbative and non-perturbative values.   
Indeed, our results for $\kappa_{\rm 1-loop}^{\rm dr}$ are
significally closer to the non-perturbative evaluations, as shown in 
Table \ref{tab14}; in fact, the two-loop dressing procedure
introduces no further improvement to the comparison.

\bigskip\noindent
{\bf Acknowledgments: } This work is supported in part by the
Research Promotion Foundation of Cyprus (Proposal Nr: 
$\rm ENTA\Xi$/0504/11, $\rm ENI\Sigma X$/0505/45).

\begin{table}
\begin{center}
\begin{minipage}{13cm}
\vskip 1cm
\caption{Input parameters $c_0$, $c_1$, $c_3$ $(c_2=0)$ \label{tab1}}  
\begin{tabular}{lr@{}lr@{}lr@{}l}
\multicolumn{1}{c}{Action}&
\multicolumn{2}{c}{$c_0$} &
\multicolumn{2}{c}{$c_1$} &
\multicolumn{2}{c}{$c_3$} \\
\tableline \hline
Plaquette               &  1&.0             &  0&                &  0&          \\
Symanzik                &  1&.6666667       & -0&.083333         &  0&          \\
TILW, $\beta c_0=8.60$  &  2&.3168064       & -0&.151791         & -0&.0128098  \\
TILW, $\beta c_0=8.45$  &  2&.3460240       & -0&.154846         & -0&.0134070  \\
TILW, $\beta c_0=8.30$  &  2&.3869776       & -0&.159128         & -0&.0142442  \\
TILW, $\beta c_0=8.20$  &  2&.4127840       & -0&.161827         & -0&.0147710  \\
TILW, $\beta c_0=8.10$  &  2&.4465400       & -0&.165353         & -0&.0154645  \\
TILW, $\beta c_0=8.00$  &  2&.4891712       & -0&.169805         & -0&.0163414  \\
Iwasaki                 &  3&.648           & -0&.331            &  0&          \\
DBW2                    & 12&.2688          & -1&.4086           &  0&          \\
\end{tabular}
\end{minipage}
\end{center}
\end{table}

\begin{table}
\begin{center}
\begin{minipage}{15cm}
\vskip 1cm
\caption{Total contribution of one-loop diagrams
\label{tab2}}
\begin{tabular}{lr@{}lr@{}lr@{}l}
\multicolumn{1}{c}{Action}&
\multicolumn{2}{c}{$\varepsilon^{(0)}$} &
\multicolumn{2}{c}{$\varepsilon^{(1)}$} &
\multicolumn{2}{c}{$\varepsilon^{(2)}$} \\
\tableline \hline
Plaquette                &-0&.1628570582(5)       &0&.0434830339(1)        &0&.01809576875(4)  \\
Symanzik                 &-0&.12805490528(8)      &0&.0378314931(2)        &0&.01476335801(5)  \\
TILW $(8.60)$           &-0&.10821568768(4)      &0&.03408560232(6)       &0&.01265991972(4)  \\
TILW $(8.45)$           &-0&.10749185625(3)      &0&.0339409375(1)        &0&.01258108895(1)  \\
TILW $(8.30)$           &-0&.1064962872(3)       &0&.0337409869(2)        &0&.012472434543(4)  \\
TILW $(8.20)$           &-0&.1058799831(2)       &0&.0336166372(1)        &0&.0124050416(1)  \\
TILW $(8.10)$           &-0&.1050866191(1)       &0&.03345591621(5)       &0&.012318127134(5)  \\
TILW $(8.00)$           &-0&.10410447893(3)      &0&.03325593631(8)       &0&.012210297749(7)  \\
Iwasaki                  &-0&.08255435613(4)      &0&.0285451387(1)        &0&.00983490867(5)  \\
DBW2                     &-0&.0364526623(2)       &0&.01581657412(5)       &0&.004280099253(2)  \\
\end{tabular}
\end{minipage}
\end{center}
\end{table}

\begin{table}
\begin{center}
\begin{minipage}{17cm}
\caption{Total contribution of two-loop diagrams of order ${\mathcal{O}}(N^2,\,c_2^0)$
\label{tab3}}
\begin{tabular}{lr@{}lr@{}lr@{}lr@{}lr@{}l}
\multicolumn{1}{c}{Action}&
\multicolumn{2}{c}{$e^{(0,2,0)}$} &
\multicolumn{2}{c}{$e^{(1,2,0)}$} &
\multicolumn{2}{c}{$e^{(2,2,0)}$} &
\multicolumn{2}{c}{$e^{(3,2,0)}$} &
\multicolumn{2}{c}{$e^{(4,2,0)}$} \\
\tableline \hline
Plaquette               &-0&.01753602(2)         &0&.00259963(2)          &-0&.000155894(8)           &-0&.000163242(2)          &-0&.00001721759(2)  \\
Symanzik                &-0&.00810366(1)         &0&.00095046(2)          &-0&.000404510(9)           &-0&.000107348(2)          &-0&.00001275904(1)  \\
TILW $(8.60)$          &-0&.00437013(7)         &0&.00019403(5)          &-0&.00045894(1)            &-0&.000078117(3)          &-0&.00001020820(1)  \\
TILW $(8.45)$          &-0&.00425575(7)         &0&.00016978(6)          &-0&.00045962(1)            &-0&.000077102(3)          &-0&.00001011451(1)  \\
TILW $(8.30)$          &-0&.00410086(7)         &0&.00013682(7)          &-0&.00046040(1)            &-0&.000075713(3)          &-0&.00000998564(1)  \\
TILW $(8.20)$          &-0&.00400636(6)         &0&.00011666(8)          &-0&.00046080(1)            &-0&.000074857(3)          &-0&.00000990584(1)  \\
TILW $(8.10)$          &-0&.00388630(6)         &0&.00009097(9)          &-0&.00046123(1)            &-0&.000073760(3)          &-0&.00000980314(1)  \\
TILW $(8.00)$          &-0&.00374009(6)         &0&.00005958(9)          &-0&.000461601(9)           &-0&.000072410(3)          &-0&.00000967600(1)  \\
Iwasaki                 &-0&.00112957(2)         &-0&.00052964(6)         &-0&.000436966(5)           &-0&.000045009(3)          &-0&.00000682353(1)  \\
DBW2                    &0&.0008481(2)           &-0&.00085301(8)         &-0&.00018540(1)            &-0&.000006164(3)          &-0&.00000173502(3)  \\
\end{tabular}
\vskip 1cm
\caption{Total contribution of two-loop diagrams of order ${\mathcal{O}}(N^0,\,c_2^0)$
\label{tab4}}
\begin{tabular}{lr@{}lr@{}lr@{}lr@{}lr@{}l}
\multicolumn{1}{c}{Action}&
\multicolumn{2}{c}{$e^{(0,0,0)}$} &
\multicolumn{2}{c}{$e^{(1,0,0)}$} &
\multicolumn{2}{c}{$e^{(2,0,0)}$} &
\multicolumn{2}{c}{$e^{(3,0,0)}$} &
\multicolumn{2}{c}{$e^{(4,0,0)}$} \\
\tableline \hline
Plaquette               &0&.01656633(2)        &-0&.00055904(1)        &0&.002622771(7)         &0&.000158125(2)         &0&.00004282674(2)  \\
Symanzik                &0&.00605656(1)        &0&.000935801(6)        &0&.002120980(9)         &0&.000104973(2)         &0&.00002971553(1)  \\
TILW $(8.60)$          &0&.00202637(3)        &0&.00157890(3)         &0&.001790242(9)         &0&.000076167(2)         &0&.00002260669(1)  \\
TILW $(8.45)$          &0&.00190729(3)        &0&.00159800(3)         &0&.001777415(9)         &0&.000075164(3)         &0&.00002235603(1)  \\
TILW $(8.30)$          &0&.00174666(3)        &0&.00162375(2)         &0&.001759689(9)         &0&.000073791(3)         &0&.00002201243(1)  \\
TILW $(8.20)$          &0&.00164901(3)        &0&.00163939(2)         &0&.001748661(9)         &0&.000072944(3)         &0&.00002180041(1)  \\
TILW $(8.10)$          &0&.00152532(3)        &0&.00165917(2)         &0&.001734421(9)         &0&.000071859(3)         &0&.00002152826(1)  \\
TILW $(8.00)$          &0&.00137535(4)    &0&.00168310(3)         &0&.00171671(1)          &0&.000070522(3)         &0&.00002119259(1)  \\
Iwasaki                 &-0&.00103022(1)       &0&.00203254(1)         &0&.001313076(3)         &0&.000043949(3)         &0&.00001423324(1)  \\
DBW2                    &-0&.0018961(2)        &0&.0016130(3)          &0&.000413397(9)         &0&.000005057(3)         &0&.00000307480(3)  \\
\end{tabular}
\vskip 1cm
\caption{Total contribution of two-loop diagrams containing closed fermion loops
\label{tab5}}
\begin{tabular}{lr@{}lr@{}lr@{}lr@{}lr@{}l}
\multicolumn{1}{c}{Action}&
\multicolumn{2}{c}{$\tilde{e}^{(0)}$} &
\multicolumn{2}{c}{$\tilde{e}^{(1)}$} &
\multicolumn{2}{c}{$\tilde{e}^{(2)}$} &
\multicolumn{2}{c}{$\tilde{e}^{(3)}$} &
\multicolumn{2}{c}{$\tilde{e}^{(4)}$} \\
\tableline \hline
Plaquette               &0&.00118621(2)         &-0&.000546197(8)       &0&.001365146(9)        &-0&.000692228(3)   &-0&.00019809791(7)  \\
Symanzik                &0&.00081496(1)         &-0&.000448276(6)       &0&.001041379(8)        &-0&.000574521(3)   &-0&.0001453370(2)  \\
TILW $(8.60)$          &0&.00063643(1)         &-0&.000389464(5)       &0&.000857737(3)        &-0&.000500011(5)   &-0&.0001148491(1)  \\
TILW $(8.45)$          &0&.00063033(1)         &-0&.000387269(5)       &0&.000851127(3)        &-0&.000497194(5)   &-0&.0001137544(1)  \\
TILW $(8.30)$          &0&.00062198(1)         &-0&.000384243(5)       &0&.000842047(3)        &-0&.000493307(5)   &-0&.0001122515(1)  \\
TILW $(8.20)$          &0&.00061684(1)         &-0&.000382366(5)       &0&.000836433(3)        &-0&.000490894(5)   &-0&.0001113227(1)  \\
TILW $(8.10)$          &0&.00061025(1)         &-0&.000379946(5)       &0&.000829214(4)        &-0&.000487781(4)   &-0&.0001101288(1)  \\
TILW $(8.00)$          &0&.00060214(1)         &-0&.000376945(5)       &0&.000820289(4)        &-0&.000483915(4)   &-0&.0001086536(1)  \\
Iwasaki                 &0&.00043546(1)         &-0&.00030800(1)        &0&.000629274(8)        &-0&.000395294(3)   &-0&.0000779538(3)  \\
DBW2                    &0&.00015833(3)         &-0&.00014883(4)        &0&.00024756(2)         &-0&.00018242(5)        &-0&.0000213595(6)  \\
\end{tabular}
\end{minipage}
\end{center}
\end{table}

\begin{table}
\begin{center}
\begin{minipage}{15.4cm}
\caption{Total contribution of two-loop diagrams containing the parameter $c_2$ (part 1)
\label{tab6}}
\begin{tabular}{lr@{}lr@{}lr@{}lr@{}lr@{}l}
\multicolumn{1}{c}{Action}&
\multicolumn{2}{c}{$e^{(0,0,1)}$} &
\multicolumn{2}{c}{$e^{(1,0,1)}$} &
\multicolumn{2}{c}{$e^{(2,0,1)}$} &
\multicolumn{2}{c}{$e^{(0,2,1)}$} &
\multicolumn{2}{c}{$e^{(1,2,1)}$} \\
\tableline \hline
Plaquette               &0&.077167(3)           &-0&.019808(3)          &-0&.0085415(2)         &-0&.047102(4)          &0&.010439(3)  \\
Symanzik                &0&.034929(2)           &-0&.010895(2)          &-0&.0041454(2)         &-0&.017940(2)          &0&.004491(2)  \\
TILW $(8.60)$          &0&.020247(1)           &-0&.007117(2)          &-0&.0024559(1)         &-0&.008702(1)          &0&.002251(1)  \\
TILW $(8.45)$          &0&.019816(1)           &-0&.006998(2)          &-0&.0024050(1)         &-0&.008448(1)          &0&.002185(1)  \\
TILW $(8.30)$          &0&.019235(1)           &-0&.006835(2)          &-0&.0023362(1)         &-0&.0081078(6)         &0&.0020973(9)  \\
TILW $(8.20)$          &0&.018881(1)           &-0&.006736(2)          &-0&.0022942(1)         &-0&.0079023(7)         &0&.002044(1)  \\
TILW $(8.10)$          &0&.018433(1)           &-0&.006609(2)          &-0&.0022410(1)         &-0&.0076431(9)         &0&.0019761(8)  \\
TILW $(8.00)$          &0&.017888(1)           &-0&.006454(2)          &-0&.0021762(1)         &-0&.0073300(6)         &0&.0018940(6)  \\
Iwasaki                 &0&.0087615(7)          &-0&.003656(1)          &-0&.00107856(8)        &-0&.0027484(4)         &0&.0006646(5)  \\
DBW2                    &0&.0007907(2)          &-0&.0004889(3)         &-0&.00008343(2)        &0&.0001308(2)          &-0&.0001587(3)  \\
\end{tabular}
\end{minipage}
%
\phantom{a} \hskip -0.8cm 
\begin{minipage}{17cm}
\vskip 7mm
\caption{Total contribution of two-loop diagrams containing the parameter $c_2$ (part 2)
\label{tab7}}
\begin{tabular}{lr@{}lr@{}lr@{}lr@{}lr@{}l}
\multicolumn{1}{c}{Action}&
\multicolumn{2}{c}{$e^{(2,2,1)}$} &
\multicolumn{2}{c}{$e^{(3,2,1)}$} &
\multicolumn{2}{c}{$e^{(0,2,2)}$} &
\multicolumn{2}{c}{$e^{(1,2,2)}$} &
\multicolumn{2}{c}{$e^{(2,2,2)}$} \\
\tableline \hline
Plaquette               &0&.0039245(3)           &-0&.0000842143(1)       &-0&.09448252(9)        &0&.02755993(3)         &0&.010521016(1)  \\
Symanzik                &0&.0014622(1)           &-0&.0000454986(1)       &-0&.03417549(2)        &0&.01248953(1)         &0&.0041047891(2)  \\
TILW $(8.60)$          &0&.0006472(1)           &-0&.00002872341(6)      &-0&.017374635(6)       &0&.007205477(3)        &0&.0021218443(2)  \\
TILW $(8.45)$          &0&.0006251(1)           &-0&.00002818123(6)      &-0&.016917713(6)       &0&.007049188(2)        &0&.0020666192(2)  \\
TILW $(8.30)$          &0&.0005954(1)           &-0&.00002744385(5)      &-0&.016304614(5)       &0&.006838088(3)        &0&.0019924047(3)  \\
TILW $(8.20)$          &0&.0005775(1)           &-0&.00002699223(5)      &-0&.015933835(5)       &0&.006709626(4)        &0&.0019474604(2)  \\
TILW $(8.10)$          &0&.0005550(1)           &-0&.00002641646(5)      &-0&.015466270(5)       &0&.006546741(4)        &0&.0018907121(3)  \\
TILW $(8.00)$          &0&.0005279(1)           &-0&.00002571231(5)      &-0&.014902324(4)       &0&.006348924(5)        &0&.0018221643(3)  \\
Iwasaki                 &0&.00015719(6)          &-0&.00001249281(2)      &-0&.00596123(2)        &0&.00295502(1)         &0&.0007286816(4)  \\
DBW2                    &-0&.00002436(1)         &-0&.00000050404(9)      &-0&.00028731(2)        &0&.00020317(4)         &0&.0000278810(8)  \\
\end{tabular}
\end{minipage}
\begin{minipage}{15cm}
\vskip 7mm
\caption{Contribution of one-loop diagrams, for the Iwasaki action
\label{tab8}}
\begin{tabular}{cr@{}lr@{}l}
\multicolumn{1}{c}{$i$}&
\multicolumn{2}{c}{$\varepsilon^{(i)}_1$} &
\multicolumn{2}{c}{$\varepsilon^{(i)}_2$} \\
\tableline \hline
0         &-0&.05602636832(2)       & -0&.02652798781(3) \\
1         &0&                       &  0&.0285451387(1) \\
2         &0&                       &  0&.00983490867(5) \\
\end{tabular}
\end{minipage}
\begin{minipage}{15cm}
\vskip 7mm
\caption{Contribution of diagrams 3, 4, 6, for the Iwasaki action
\label{tab9}}
\begin{tabular}{cccr@{}lr@{}lr@{}l}
\multicolumn{1}{c}{$i$}&
\multicolumn{1}{c}{$j$}&
\multicolumn{1}{c}{$k$}&
\multicolumn{2}{c}{$e^{(i,j,k)}_{3}$} &
\multicolumn{2}{c}{$e^{(i,j,k)}_{4}$} &
\multicolumn{2}{c}{$e^{(i,j,k)}_{6}$} \\
\tableline \hline
0 & 0 & 0     &-0&.0003923686(9)      &-0&.000743134(3)       &-0&.0000714882(8) \\
0 & 2 & 0     &0&.0002615791(6)       &0&.000495422(2)        & 0&.0000357441(4) \\
1 & 0 & 0     &0&                     &0&.001900337(2)        & 0&               \\
1 & 2 & 0     &0&                     &0&.0017774410(9)       & 0&               \\
2 & 0 & 0     &0&                     &-0&.0010339720(2)      & 0&               \\
2 & 2 & 0     &0&                     &-0&.001041123(1)       & 0&.0002799238(4) \\
\end{tabular}
\end{minipage}
\end{center}
\end{table}

\begin{table}
\begin{center}
\begin{minipage}{15cm}
\caption{Contribution of diagrams 7-11, 14-18, 24, 26, for the Iwasaki action
\label{tab10}}
\begin{tabular}{cccr@{}lr@{}lr@{}lr@{}l}
\multicolumn{1}{c}{$i$}&
\multicolumn{1}{c}{$j$}&
\multicolumn{1}{c}{$k$}&
\multicolumn{2}{c}{$e^{(i,j,k)}_{7-11}$} &
\multicolumn{2}{c}{$e^{(i,j,k)}_{14-18}$} &
\multicolumn{2}{c}{$e^{(i,j,k)}_{24}$} &
\multicolumn{2}{c}{$e^{(i,j,k)}_{26}$} \\
\tableline \hline
0 & 0 & 0  &0&.00042802(1)          &-0&.000195263(2)        &0&                     &0&                 \\
0 & 0 & 1  &0&.0057103(7)           &0&.0030512(2)           &0&                     &0&                 \\
0 & 2 & 0  &-0&.00111995(2)         &-0&.00029748(1)         &0&                     &-0&.000298742(2)   \\
0 & 2 & 1  &-0&.0022472(3)          &-0&.0008718(2)          &0&                     &0&.0003705893(7)   \\
0 & 2 & 2  &-0&.00371263(2)         &-0&.00224859(1)         &0&                     &0&                 \\
1 & 0 & 0  &0&                      &0&.00064534(1)          &0&                     &0&                 \\
1 & 0 & 1  &0&                      &-0&.003656(1)           &0&                     &0&                 \\
1 & 2 & 0  &0&                      &0&.00011079(6)          &-0&.000144897(2)       &0&.000429899(1)    \\
1 & 2 & 1  &0&                      &0&.0006450(5)           &0&.000248682(4)        &-0&.00022905(1)    \\
1 & 2 & 2  &0&                      &0&.00295502(1)          &0&                     &0&                 \\
2 & 0 & 0  &0&                      &-0&.000000974(1)        &0&                     &0&                 \\
2 & 0 & 1  &0&                      &-0&.00107856(8)         &0&                     &0&                 \\
2 & 2 & 0  &0&                      &0&.000141960(3)         &0&.000042314(2)        &0&.0003303085(7)   \\
2 & 2 & 1  &0&                      &0&.00039546(6)          &0&.00002909398(7)      &-0&.000267364(2)   \\
2 & 2 & 2  &0&                      &0&.0007286816(4)        &0&                     &0&                 \\
3 & 2 & 0  &0&                      &0&                      &0&                     &-0&.000019835(1)   \\
3 & 2 & 1  &0&                      &0&                      &0&                     &-0&.00001249281(2) \\
\end{tabular}
\end{minipage}
\begin{minipage}{15cm}
\vskip 1cm
\caption{Contribution of diagrams 12, 13, 19, 20, for the Iwasaki action
\label{tab11}}
\begin{tabular}{cr@{}lr@{}l}
\multicolumn{1}{c}{$i$}&
\multicolumn{2}{c}{$\tilde{e}^{(i)}_{12-13}$} &
\multicolumn{2}{c}{$\tilde{e}^{(i)}_{19-20}$} \\
\tableline \hline
0         &0&.000261920(6)         &0&.000173538(9)    \\
1         &-0&.0000308339(1)       &-0&.00027717(1)   \\
2         &0&.000370942(2)         &0&.000258332(8)    \\
3         &0&                      &-0&.000395294(3)   \\
4         &0&                      &-0&.0000779538(3) \\
\end{tabular}
\end{minipage}
\begin{minipage}{15cm}
\vskip 1cm
\caption{Contribution of diagrams 21-23, 25, 27, 28, for the Iwasaki action
\label{tab12}}
\begin{tabular}{cccr@{}lr@{}lr@{}lr@{}l}
\multicolumn{1}{c}{$i$}&
\multicolumn{1}{c}{$j$}&
\multicolumn{1}{c}{$k$}&
\multicolumn{2}{c}{$e^{(i,j,k)}_{21-23}$} &
\multicolumn{2}{c}{$e^{(i,j,k)}_{25}$} &
\multicolumn{2}{c}{$e^{(i,j,k)}_{27}$} &
\multicolumn{2}{c}{$e^{(i,j,k)}_{28}$} \\
\tableline \hline
0 & 0 & 0     &0&.000373419(3)          &-0&.000158621(4)        &-0&.000094848(3)         &-0&.0001759336(5) \\
0 & 2 & 0     &-0&.000373419(3)         &0&.000079311(2)         &0&                       &0&.0000879668(3)  \\
1 & 0 & 0     &-0&.000887295(1)         &0&.0001396819(4)        &0&.000045158(4)          &0&.0001893113(5)  \\
1 & 2 & 0     &0&.000887295(1)          &0&.000085189(2)         &0&                       &-0&.000120480(1)  \\
2 & 0 & 0     &0&.000194437(1)          &-0&.0000319392(3)       &0&.000168506(2)          &-0&.0000509266(2) \\
2 & 2 & 0     &-0&.000194437(1)         &-0&.000005787(2)        &0&                       &0&.0000098758(1)  \\
3 & 0 & 0     &0&.000059183(3)          &0&                      &-0&.000015234(1)         &0&                \\
3 & 2 & 0     &-0&.000059183(3)         &0&.0000172022(5)        &0&                       &0&.0000168072(6)  \\
4 & 0 & 0     &0&.00000682353(1)        &0&                      &0&.000007409712(6)       &0&                \\
4 & 2 & 0     &-0&.00000682353(1)       &0&                      &0&                       &0&                \\
\end{tabular}
\end{minipage}
\end{center}
\end{table}

\begin{table}
\begin{center}
\begin{minipage}{15cm}
\caption{Results for $dm^{\rm dr}_{\rm (1-loop)}$
(Eq. (\ref{dm1loopDressed})), with $N=3$
\label{tab13}}
\begin{tabular}{llr@{}lr@{}lr@{}l}
\multicolumn{1}{c}{Action}&
\multicolumn{1}{c}{$\beta$}&
\multicolumn{2}{c}{$\varepsilon^{(0)}_{dr}$} &
\multicolumn{2}{c}{$\varepsilon^{(1)}_{dr}$} &
\multicolumn{2}{c}{$\varepsilon^{(2)}_{dr}$} \\
\tableline \hline
Plaquette      &6.00            &-0&.579221119(2)         &0&.1159547570(3)         &0&.03618067788(9)  \\
Symanzik       &5.00            &-0&.4869797578(8)        &0&.1121369999(4)         &0&.03538605357(4)  \\
Symanzik       &5.07            &-0&.478756110(2)         &0&.11072412996(5)        &0&.03507238306(5)  \\
Symanzik       &6.00            &-0&.3915226522(2)        &0&.0947962001(5)         &0&.03124138429(9)  \\
TILW $(8.60)$  &3.7120          &-0&.5358770348(7)        &0&.1265917638(3)         &0&.03813963851(4)  \\
TILW $(8.45)$  &3.6018          &-0&.5497415338(3)        &0&.1291104644(3)         &0&.0386337113(1)  \\
TILW $(8.30)$  &3.4772          &-0&.5651407386(9)        &0&.1319263769(1)         &0&.0391695069(1)  \\
TILW $(8.20)$  &3.3985          &-0&.5756111531(9)        &0&.1337937558(7)         &0&.03951713046(7)  \\
TILW $(8.10)$  &3.3107          &-0&.5870122772(4)        &0&.1358437825(6)         &0&.0398899143(3)  \\
TILW $(8.00)$  &3.2139          &-0&.599415804(1)         &0&.138085996(2)          &0&.0402877133(4)  \\
Iwasaki        &1.95            &-0&.757856451(1)         &0&.1671007819(8)         &0&.044746728234(1)  \\
Iwasaki        &2.20            &-0&.6555102085(5)        &0&.1537748193(6)         &0&.04293183656(3)  \\
Iwasaki        &2.60            &-0&.541348980(1)         &0&.1359882440(3)         &0&.03967626495(6)  \\
DBW2           &0.6508          &-0&.7749943512(7)        &0&.1847244889(1)         &0&.04731717866(3)  \\
DBW2           &0.8700          &-0&.574781578(1)         &0&.1575688409(9)         &0&.04281261980(1)  \\
DBW2           &1.0400          &-0&.4822863343(9)        &0&.1412499230(5)         &0&.039186543574(5)  \\
\end{tabular}
\end{minipage}
\end{center}
\end{table}

\begin{table}
\begin{center}
\begin{minipage}{16cm}
\begin{center}
\caption{One- and two-loop results, and non-perturbative estimates for
  $\kappa_c$
\label{tab14}}
\begin{tabular}{lccr@{}lr@{}lr@{}lr@{}lr@{}lc}
\multicolumn{1}{c}{Action}&
\multicolumn{1}{c}{$N_f$}&
\multicolumn{1}{c}{$\beta$}&
\multicolumn{2}{c}{$c_{\rm SW}$}&
\multicolumn{2}{c}{$\kappa_{\rm 1-loop}$} &
\multicolumn{2}{c}{$\kappa_{\rm 2-loop}$} &
\multicolumn{2}{c}{$\kappa_{\rm 1-loop}^{\rm dr}$} &
\multicolumn{2}{c}{$\kappa_{\rm 2-loop}^{\rm dr}$} &
\multicolumn{1}{c}{$\kappa_c^{\rm non-pert}$\ [Ref.]} \\
\tableline \hline
Plaquette & 0 & 6.00   & 1&.479  & 0&.1301 & 0&.1335 & 0&.1362 & 0&.1362 & 0.1392  \hskip 3mm \cite{Bowler}\\
Plaquette & 0 & 6.00   & 1&.769  & 0&.1275 & 0&.1306 & 0&.1337 & 0&.1332 & 0.1352  \hskip 3mm \cite{Luscher1996} \\[1ex]
Plaquette & 2 & 5.29   & 1&.9192 & 0&.1262 & 0&.1307 & 0&.1353 & 0&.1341
& $\displaystyle{0.1373 \atop 0.1363}$ $\displaystyle{\cite{UKQCD} \atop \cite{Sommer}}$ \\[2ex]
Iwasaki   & 2 & 1.95  & 1&.53   & 0&.1292   & 0&.1368 & 0&.1388 & 0&.1379 & 0.1421  \hskip 3mm \cite{Khan} \\
TILW $(8.60)$      & 0 & 3.7120 & 1&.0      & 0&.1339 & 0&.1370 & 0&.1378 & 0&.1384 &  \\
TILW $(8.00)$      & 0 & 3.2139 & 1&.0      & 0&.1348 & 0&.1387 & 0&.1397 & 0&.1406 &  \\
DBW2               & 2 & 0.87   & 0&.0      & 0&.1502 & 0&.1384 & 0&.1460 & 0&.1479 &  \\
DBW2               & 2 & 0.87   & 1&.0      & 0&.1352 & 0&.1372 & 0&.1379 & 0&.1379 &  \\
DBW2               & 2 & 1.04   & 0&.0      & 0&.1454 & 0&.1375 & 0&.1421 & 0&.1434 &  \\
DBW2               & 2 & 1.04   & 1&.0      & 0&.1334 & 0&.1348 & 0&.1352 & 0&.1352 &  \\
\end{tabular}
\end{center}
\end{minipage}
\end{center}
\end{table}


\end{document}